\documentclass[twoside]{article}
\usepackage{frascatiphys-in_R}
\usepackage{graphicx} 
\usepackage{hyperref}
\newcommand{\hhref}[1]{\href{http://#1}{#1}}

\catcode`\@=11
\def\ps@ppt{\def\@oddhead{\qquad \textsc{Comunicare Fisica 
all'americana}\hfil \thepage\qquad}
\def\@evenhead{\qquad\thepage \hfil \textsc{Chris Quigg} \qquad}
\def\@oddfoot{}\def\@evenfoot{}}    
\catcode`\@=12

\begin{document}
\title{ 
COMUNICARE FISICA ALL'AMERICANA \\
}
\author{
Chris Quigg\thanks{~~Electronic mail: quigg@fnal.gov \hfill 
\fbox{\textsf{FERMILAB--CONF--05/490--T}}}      \\
{\em Fermi National Accelerator Laboratory} \\ {\em P.O. Box 500, Batavia, 
Illinois USA 60510} 
}
\maketitle
\baselineskip=11.6pt
\begin{abstract}
    I survey motivations for education and outreach initiatives in the
    American context and explore the value of communicating physics for
    physicists and for the wider society.  I describe the roles of
    large institutions, professional organizations, and funding
    agencies and cite some individual actions, local activities, and
    coordinated national programs.  I note the emergence of
    transnational enterprises---not only to carry out research, but to
    communicate physics. A brief appendix collects some useful internet 
    resources. 
\end{abstract}
\baselineskip=14pt
\section{Introductions}
It is my great pleasure to join Italian colleagues for this ambitious
workshop on communicating physics.  In my main career as a theoretical
physicist, my research has addressed many topics in particle physics.
My close engagement with experiment has made the Fermi National
 Accelerator Laboratory a stimulating scientific home for many years.
Beyond my personal activities to bring the ideas and aspirations
of particle physics to a broad audience, I have held a number of
positions in which I have been able to encourage---and take pleasure
in---the outreach and education efforts of many of my colleagues.  I
have served as Head of the Fermilab Theory Group, as Deputy Director of
the Superconducting Super Collider Central Design Group, and as Chair
of the American Physical Society's Division of Particles and Fields.
In the last capacity, I oversaw the planning and execution of
\textit{Snowmass 2001:} a summer study on the future of particle
physics (see \hhref{snowmass2001.org}). I am also a member of the 
board of trustees of the Illinois Mathematics and Science Academy, a 
state-supported residential high school.

Fermilab is a U.S.\ Department of Energy national laboratory for particle
physics, operated by a consortium of 90 research universities
in the United States, Canada, Italy, and Japan.  The laboratory and its
research community are highly international; our lab has much in common
with CERN. Fermilab has some two thousand employees, an annual budget
of \$300M, and welcomes 2250 experimental ``users'' from
around the world.  Our principal research instrument is the Tevatron, a
superconducting proton synchrotron $2\pi$ kilometers in circumference
that accelerates protons and antiprotons to nearly 1~TeV. When operated
as a proton-antiproton collider, the Tevatron is the world's most
powerful microscope, giving us access to \textit{nanonanophysics} on a scale a 
billion times smaller than an atom.

Many of you have visited my country, or gained an impression of it from
Puccini's opera, \textit{La Fanciulla del West,} or from the cult 
classic film, \textit{The Blues Brothers.} Nevertheless, a few remarks 
about the country's character may provide useful context for today's 
exploration.

The United States is large in both land area and population, and much
of its area is lightly settled.  It is a young nation, at least in its
psyche, and it is the wealthiest country in the history of the planet.
The U.S.\ was founded as a nation of immigrants, and the flux of people
from diverse origins to our shores continues today.  A great degree of
social mobility is perhaps correlated with a restrained esteem for
authorities (including professors).  At the same time, unease with the
pace of change leads many to a powerful reliance on Authority.

The U.S.\ is a deeply non-Napoleonic society, characterized by much local
autonomy, notably in the responsibility for schools.  Our university
system is highly heterogeneous.  The national government assumes 
responsibility only for  a few military academies, and our most celebrated
universities are largely private institutions.  In America's heartland,
the state universities serve as the leading cultural institutions.
Though academic science is quite young in America, research has
traditionally been centered in universities---research and teaching go
hand in hand---and that has led to a superb standard of postgraduate
education.

In contrast to the European standard, we do not have dominant national
newspapers.  Our vast radio and television empires, which seem
increasingly preoccupied with gossip and celebrity trials, evince
little interest in science.  National Public Radio, especially in its
local programming, is receptive to science and culture.  All of these
factors---beginning with the vastness of the country---enhance the
importance of local outreach and education efforts.
\section{Incentives for Education \& Outreach Efforts}%
Some of the motivations I would cite for physics communication are
similar to those evoked by earlier speakers.  First comes the teaching
imperative: the desire not only to pass on information about the minute
particulars of current research, but also to cultivate an understanding
of the scientific worldview, with its rejection of Authority, reliance
on experiment, and celebration of doubt.  Next comes the propaganda
function, which I understand here in a positive light:
\textit{propaganda fide} for our attachment to reason, observation, and
controlled experiment; the development of what social scientists call
``diffuse good will'' toward the scientific enterprise; and an attempt 
to influence behavior and gain acceptance and support for our 
undertakings.

Since we scientists devote our lives to exploration, I see outreach and
education as a mode of exploring the world---an opportunity to learn
from and about others.  Some of my colleagues emphasize the value of
``inreach''---to see ourselves as others see us!  Previous speakers
have spoken of engaged learning; we might well aspire to
\textit{engaged teaching.} Participation in outreach programs can open
our own students to the wider world.  Our communications efforts can
also help to change the face of physics---to attract women in greater
numbers and to show minority and new immigrant groups the appeal of
careers in and around science.  Finally, outreach and education
programs are good for physicists: the joy that comes from sharing our
numinous adventure with others is marvelous psychotherapy!

\section{U.S. Organizations and Institutions}
The American Physical Society (\hhref{www.aps.org}),
the principal professional organization for academic and industrial
physicists in the United States, offers many education and outreach
programs (\hhref{www.aps.org/educ/}).  Its
\textit{Physics Central} web site
(\hhref{physicscentral.org}) aims to
communicate the excitement and importance of physics to everyone.
\textit{Physical Review Focus} (\hhref{focus.aps.org}) stories explain
selected physics research published in the APS journals
\textit{Physical Review} and \textit{Physical Review Letters.} Several
of the Divisions of the APS maintain vigorous outreach and education
programs, and the Forum on Education (\hhref{www.aps.org/units/fed/})
offers a meeting place for members interested in education.

The American Association of Physics Teachers
(\hhref{www.aapt.org}), with $11\,000$ members
around the world, aspires to be the leading organization for physics
education.  It publishes \textit{The Physics Teacher} and the
\textit{American Journal of Physics,} and strives to provide the most
current resources and up-to-date research needed to enhance a physics
educator's professional development.

The American Institute of Physics, a federation of ten Member Societies
representing the spectrum of physical sciences, supports a number of
educational efforts, described at
\hhref{www.aip.org/education/}.  The
electronic newsletter \textit{Physics News Update}
(\hhref{www.aip.org/physnews/update/})
is a digest of physics news items arising from physics meetings and
journals, newspapers and magazines, and other news sources.  The AIP
also maintains the \textit{Physics News Graphics} image archive
(\hhref{www.aip.org/png/}).

The American Association for the Advancement of Science 
(\hhref{www.aaas.org}) is the 
largest scientific organization in the United States. AAAS sponsors an 
annual meeting---in which physics plays a minor, but growing, 
part---that is a major event for science writers from much of the 
world. It also animates a great range of education programs and 
activities to promote public awareness of science 
and its role in public policy (\hhref{www.aaas.org/programs/education/}).

The Department of Energy's Office of Science
(\hhref{www.science.doe.gov}) is the largest supporter of basic
research in the physical sciences in the United States.  It is also the
main patron of large research instruments for high-energy physics,
nuclear physics, and fusion energy sciences.  The Office of Science
sponsors a range of  education initiatives through its Workforce
Development for Teachers and Scientists program
(\href{http://www.scied.science.doe.gov/scied/sci_ed.htm}{www.scied.science.doe.gov/scied/sci\_ed.htm}).
The Office of High Energy Physics points with pride to educational and
outreach efforts of its national laboratories and university-based
research groups but does not set specific expectations or 
requirements for research contractors.
The DOE and NSF jointly sponsor the \textsl{QuarkNet} program described
in \S\ref{subsec:qn}.

The National Science Foundation (\hhref{www.nsf.gov}), which mainly
funds individual investigators, is the other principal source of
support for basic research in physics.  Support for science and
engineering education, from pre-Kindergarten through graduate school
and beyond, is  essential to NSF's mission.  In addition to
judging the intellectual merit of scientific proposals to the NSF, reviewers are
asked to take into account the ``broader impacts'' of the proposed
research program.  Specifically, \textit{How well does the activity advance
discovery and understanding while promoting teaching, training, and
learning?  How well does the proposed activity broaden the
participation of underrepresented groups?} Last year alone, NSF 
grantees in particle physics reached more than $100\,000$ school-age 
children. NSF directly funds research into effective educational 
practice, and supports research participation for students and teachers.

The National Aeronautics and Space Administration
(\hhref{www.nasa.gov}) has an enviable record in engaging the public
imagination in the value of exploration and in the astronomical
sciences. NASA requires an extensive outreach and education effort 
for each of the missions it supports.

All of the major particle physics laboratories have extensive public
affairs and education offices.  Many in the \textit{Comunicare Fisica}
audience will know Judy Jackson of Fermilab (\hhref{www.fnal.gov}) and
Neil Calder of the Stanford Linear Accelerator Center
(\hhref{www.slac.stanford.edu}), and the magazine \textsf{symmetry}
(\hhref{www.symmetrymag.org}) recently launched as a joint venture of
the two institutions.  Both labs have lively programs for science
writing interns.  Together with other particle physics laboratories
around the world, U.S.\ institutions have launched
\hhref{interactions.org}, a central resource for communicators of
particle physics.  To celebrate  the World Year of Physics,
\textsf{interactions} has organized \textit{Quantum Diaries}
(\hhref{interactions.org/quantumdiaries}), a web site that follows
physicists from around the world through their blogs, photos, and
videos.

The International Linear Collider (\hhref{linearcollider.org}), 
which many of us see as the next great accelerator project after the 
Large Hadron Collider at CERN, is organized as an international 
design effort, with an international communications team.

The Kavli Institute for Theoretical Physics
(\hhref{www.itp.ucsb.edu}), on the Santa Barbara campus of the
University of California, is funded by the National Science Foundation
and the University of California.  The KITP program of workshops
encompasses research in theoretical physics very broadly understood.
The Kavli Institute has created a journalist in residence
program,\footnote{To my knowledge, no physics institution in the U.S.\
offers an experience comparable to the Woods Hole Oceanographic
Institution's Ocean Science Journalism Fellowship
(\href{http://www.whoi.edu/home/news/media_jfellowship.html}{www.whoi.edu/home/news/media\_jfellowship.html}).}
and also has an artist in residence.

The Particle Data Group (\hhref{pdg.lbl.gov}), an international
collaboration that reviews particle physics and related areas of
astrophysics, compiles and analyzes data on particle properties.  From
its U.S.\ center at Lawrence Berkeley National Laboratory, the 
PDG produces and distributes a wealth of educational materials, 
including the famous standard model wall chart described in 
\S\ref{subsec:smwc}.

The Division of Particles and Fields (\hhref{www.aps.org/units/dpf/})
and Division of Physics of Beams (\hhref{www.aps.org/units/dpb/}) of
the American Physical Society, like many of their counterparts in 
other subfields, support a range of ongoing and special programs in 
education and outreach. Increasingly, individual experiments are 
making significant efforts to bring intelligible accounts of their 
research to public notice.

\section{Examples from Particle Physics in the United States}
Now I would like to briefly describe a number of education and outreach
initiatives carried out by American particle physicists.  I have chosen
these examples out of my own experiences to illustrate the lessons I
will draw in \S\ref{sec:lessons}, including  significant
unplanned successes. See the Appendix for additional resources.
\subsection{Saturday Morning Physics at Fermilab}
When Leon Lederman became Director of Fermilab in 1979, he created the
Saturday Morning Physics program (\hhref{www-ppd.fnal.gov/smp-w/}) for
high school students curious about science.  For more than twenty-five
years, we have welcomed students---and some of their teachers---to the
lab on Saturday mornings for ten-week series of lectures (two hours)
and in-depth tours of many activities at the laboratory. The lab runs 
three sessions of SMP each year, with as many as 120 students in 
each ``class.'' By now, more than six thousand students have been 
exposed to the ideas of particle physics, the wonders of exotic 
technologies, and the spirit of scientific inquiry. 

We made an important discovery during the first year of Saturday
Morning Physics: a number of high school science teachers accompanied
their students, and devoted ten consecutive Saturday mornings to
learning about particle physics.  Though they didn't all have deep
preparation to teach physics, they all had sharp intelligence,
inquisitive minds, and high enthusiasm.  They all wanted to become
better teachers; our lab could provide information and encouragement
and---perhaps most important---the opportunity for them to meet and
support each other.  I believe that our accidental discovery of these
teachers launched the initiative that blossomed, under the inspired
guidance of Marge Bardeen and Stanka Jovanovic, into the lab's
Education Office (\hhref{www-ed.fnal.gov}) and Leon M.\ Lederman Science
Education Center.

\subsection{Physics Vans}
A number of university physics departments organize physics road shows
for schools, shopping centers, and community events.  These feature
demonstrations, often performed theatrically with an amusing twist, by
physics undergraduates and graduate students, and hands-on experiments
for audience members.  Physics vans spark curiosity and put a human
face on physics. They are also great fun for the student performers!

\subsection{Standard model wall chart \label{subsec:smwc}}
\begin{figure}[tbh]
\begin{center}
\includegraphics[width=10cm]{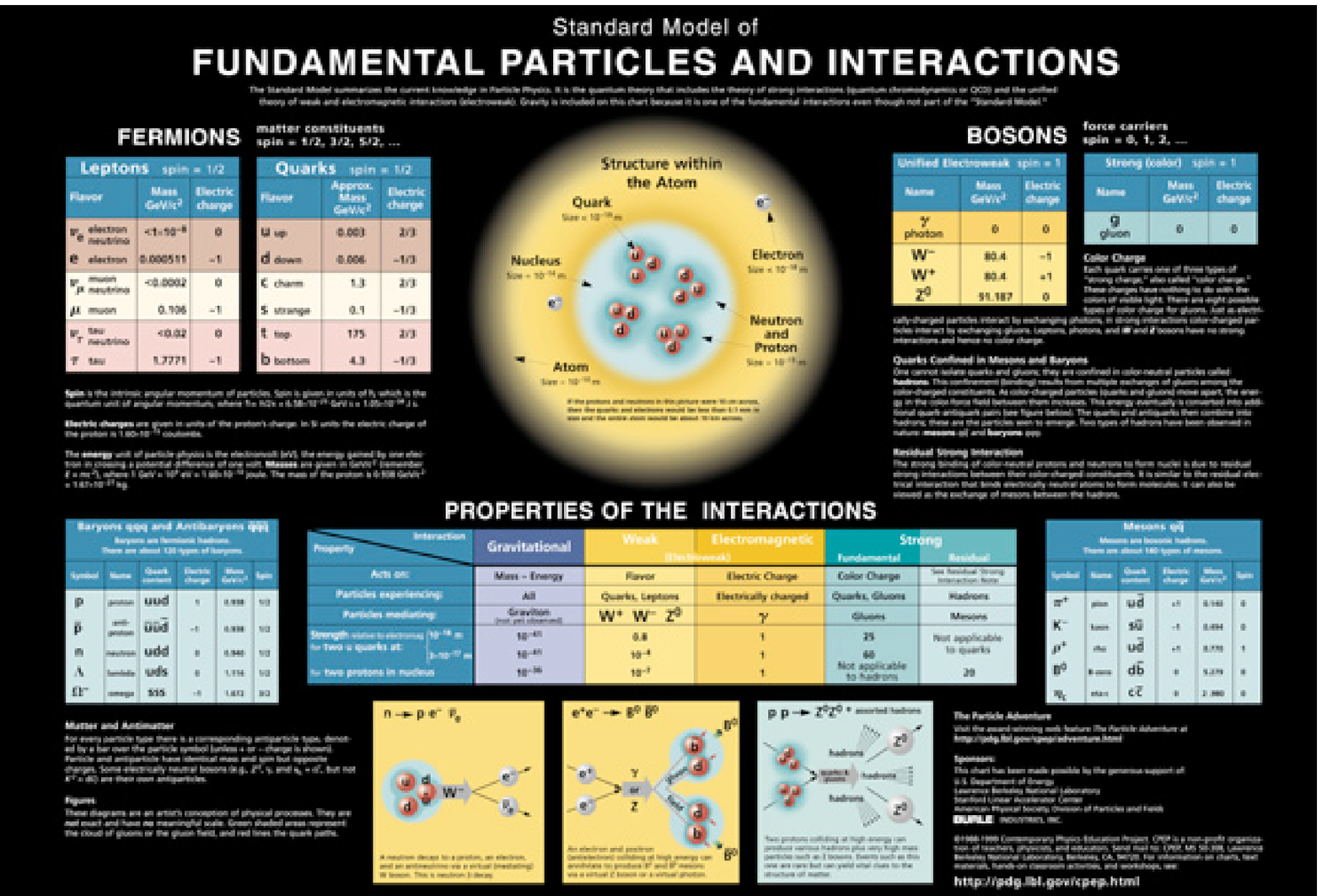}
\caption{Contemporary Physics Education Project standard-model wall chart. 
\label{fig:wallchart}}
\end{center}
\end{figure}

One of the icons of physics classrooms is the table of
particles and interactions that grew out of a Conference on Teaching
Modern Physics held at Fermilab in 1986.\cite{quandaries} The
celebrated chart (see Figure~\ref{fig:wallchart}) of fundamental
particles and interactions created by the Contemporary Physics
Education Project represents the enthusiastic work of many teachers.
Available as wall chart, poster, and place mat, this representation of
our ``standard model'' has had a global reach, with more than
$200\,000$ distributed.  It has helped move particle physics into the
classroom, and it presents many essential notions of our current
understanding. The wall chart project also stimulated \textit{The Particle 
Adventure,} an interactive web site.

\subsection{\textsl{QuarkNet} \label{subsec:qn}}
\textsl{QuarkNet} is a remarkable immersive research experience for 
high school teachers and their students that is based on lively 
ongoing partnerships 
with experimental research groups.\footnote{The program was 
established in 1998 by Michael Barnett (Lawrence Berkeley National 
Lab), Marjorie Bardeen (Fermilab), Keith Baker (Hampton University), 
and Randy Ruchti (Notre Dame), with initial funding from the National 
Science Foundation and continuing support from NSF and DOE.} Currently 
fifty-three \textsl{QuarkNet} centers operate in twenty-five states and 
Puerto Rico. They touch $100\,000$ students per year, and five 
hundred teachers and one hundred students are  research 
partners with 150 physicist mentors. One of the goals is to engage 
teachers and students in real time with data from the Large Hadron 
Collider at CERN. For now, QuarkNet centers involve twelve major 
experiments and the computing GRID.

A key tenet of the \textsl{QuarkNet} paradigm is that the teachers and students
should gain experience in assembling and commissioning real detectors.
A favorite example is the construction of cosmic-ray detectors that
consist of several paddles of plastic scintillator, photomultiplier
tubes, and the associated trigger system.  School groups are
deploying these simple detectors at schools around the country in the 
\textsl{QuarkNet} Cosmic-Ray Detector Array. Students are learning to use the 
GRID to handle calculations involving large amounts of data. (I have 
the impression that the professionals are learning how to make the 
GRID a robust tool, in the process.)

\subsection{Web lecture archives---unplanned outreach!}
\begin{figure}[tb]
   \begin{center}
       {\includegraphics[width=10cm]{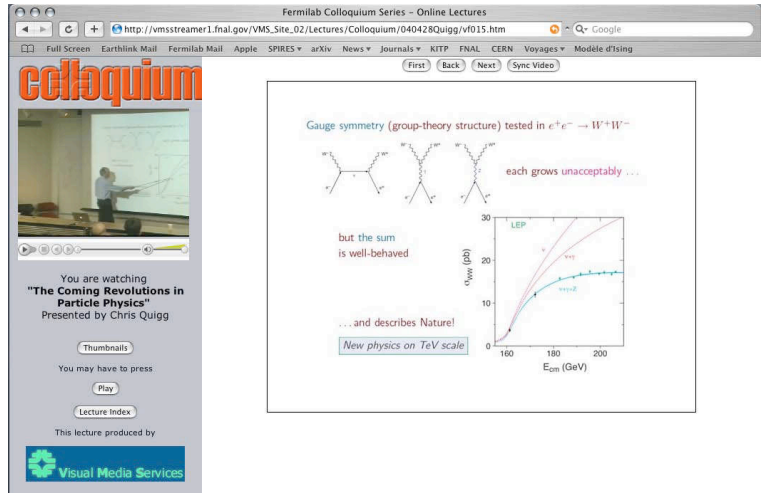}}
\caption{Appearance (in a browser) of a lecture from the Fermilab web 
archive.}
\label{fig:fish3}
\end{center}
\end{figure}

In the summer of 1999, computer scientist Chuck Severance and physicist
Steve Goldfarb, then a member of the L3 collaboration at CERN's Large
Electron-Positron Collider, tested a web lecture archive of lectures
for summer students.  I had the good fortune to be their first
experimental animal. Severance's Sync--o--matic software 
framework\cite{weblec} is simple and functional. It synchronizes a 
video stream with good-resolution images of the speaker's slides, all 
displayed in a web browser. A recent example from Fermilab's web 
lecture archive is shown in Figure~\ref{fig:fish3}.

When I showed the system to my friends in Fermilab's Visual Media
Services, they were impressed with the low resource cost (including
modest bandwidth and storage requirements) and ease of use, and were
quick to see the potential in a streaming video archive.  Today at
Fermilab, the streaming video archive boasts 1334 entries, including
colloquia, conference talks, academic training lectures, and memorable
events.\footnote{Archives of similar richness are maintained at CERN,
SLAC, and the Kavli Institute for Theoretical Physics (see the Appendix
for links).  The KITP is experimenting with the new medium of enhanced
podcasts.} We hoped that the archive might prove valuable for 
Fermilab's staff and users, as indeed it has, but didn't imagine 
that it would become part of the lab's public face---and the field's. 
In fact, many viewers from outside our community land at the video 
archive thanks to search engines, not by drilling down from the 
Fermilab home page.
\subsection{Engaging Hispanic students}
While living at Fermilab as a CDF postdoc, Aaron Dominguez (now at the
University of Nebraska) developed an educational project in Aurora,
Illinois, to improve the future and stability of the Hispanic community
by supporting the educational and social accomplishments of its young
people.  Bilingual English/Spanish Tutors (BEST) pairs high-achieving
high school student mentors with low-income elementary school children.
The BEST tutors help their younger peers with homework, reading, and
math after school twice a week.  The tutors also gain a critical sense
of responsibility for the successful education of their own Latino
community.  Aaron's program has enrolled over 60 students and 35
bilingual high school mentors; the BEST model has been replicated in
the neighboring community of Batavia.

\subsection{Tevatron postcards}
In 1997, I received an invitation to give the first Carl Sagan 
Memorial Lecture in the series, \textit{Cosmos Revisited,} at the 
Smithsonian Institution in Washington. I wanted to give members of the 
audience specimens that would stimulate them to continue the 
conversation begun by my lecture, ``The Particle Cosmos.'' With Judy 
Jackson, we conceived an edition of eight postcards depicting 
significant events---the outcome of proton-antiproton 
collisions---from the CDF and D\O\ experiments. 
\begin{figure}[h!]
   \begin{center}
       {\includegraphics[width=8cm]{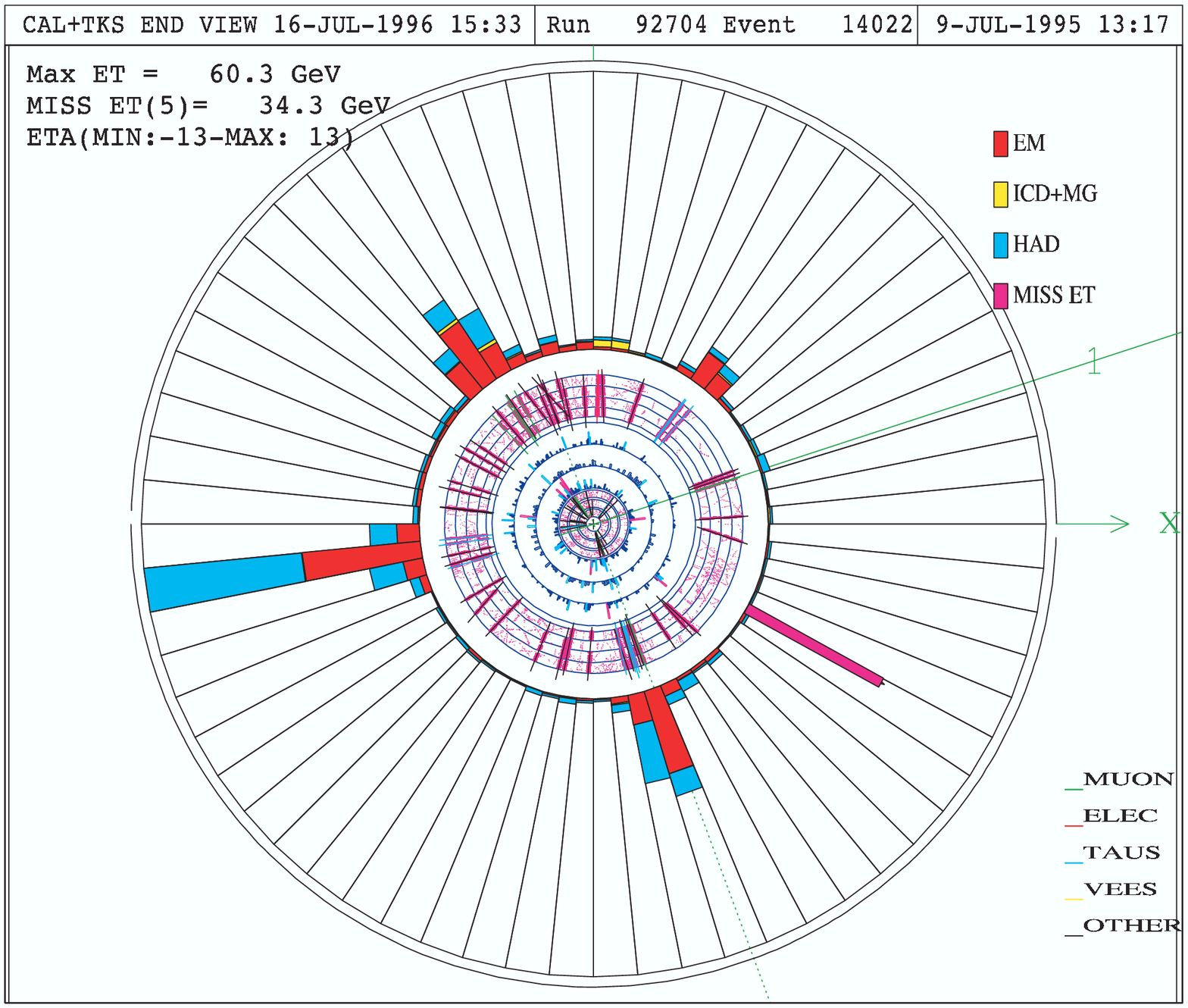}} \\
       {\includegraphics[width=8cm]{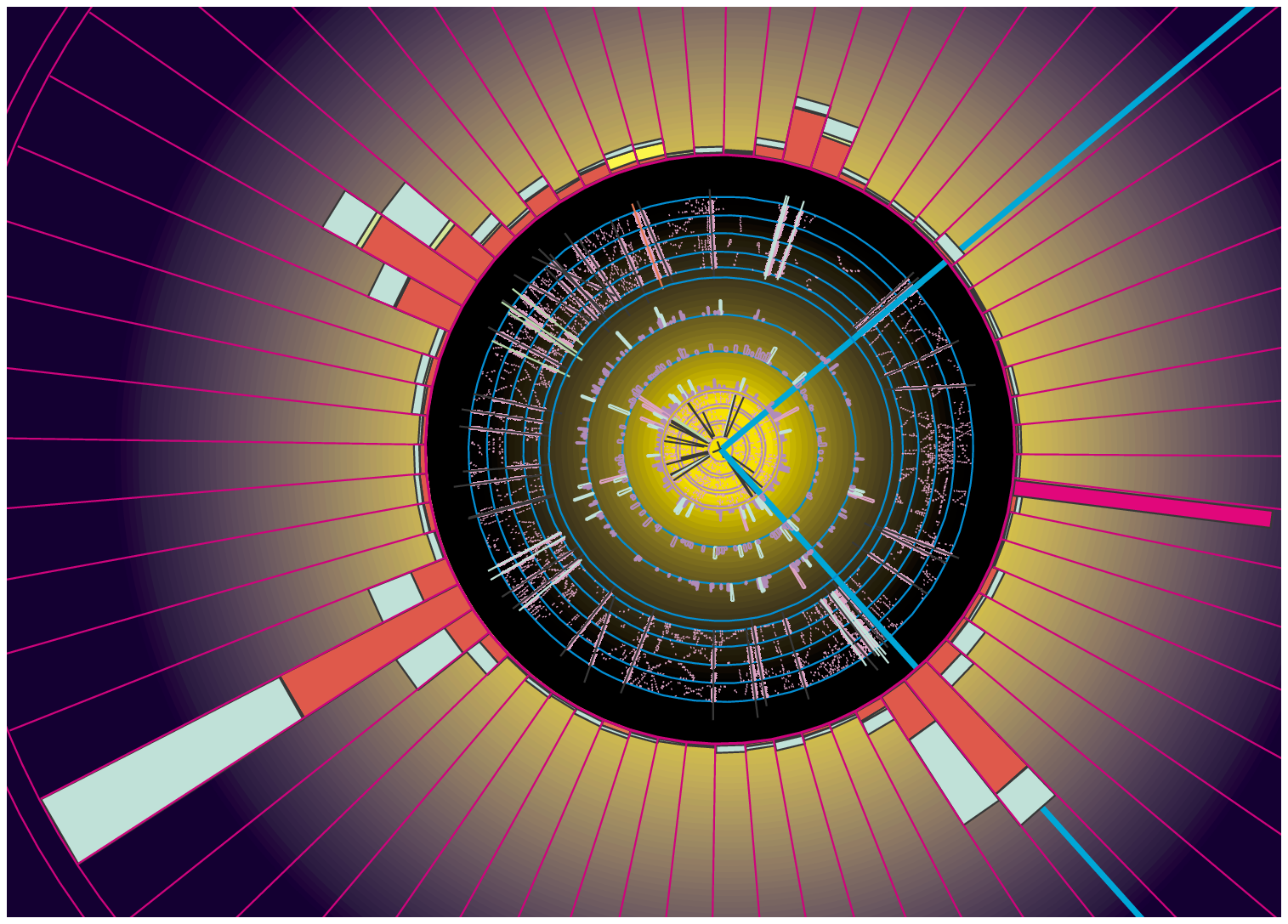}}
\caption{ A pair of top quarks reconstructed in the D\O\ experiment at
Fermilab.  This end view shows the final decay products: two muons
(turquoise), a neutrino (pink), and four jets of particles.}
\label{fig:fish2}
\end{center}
\end{figure}

We began by asking the experimental groups to submit authentic event
displays.  An example is shown in the top panel of
Figure~\ref{fig:fish2}.
This display of a top-quark event from the D\O\ experiment illustrates
an important reality: Experimenters are deeply attached to their
detectors, so the fixed detector elements are represented by strong,
assertive lines, whereas the ephemeral tracks that bear witness to
noteworthy one-time occurrences are indistinct.  The light green traces
(one solid, one dotted) representing muons---crucial markers in the
top-antitop event---are nearly invisible.

Graphic designer Bruce Kerr (\hhref{www.kerrcom.com}) preserved the
authenticity of the event displays by discreetly editing the event
display to emphasize the elements that signal top-pair production.
Except for the background sunburst that contributes visual interest and
serves as a metaphor for the conversion of energy into new forms of
matter, every element of the postcard shown in the bottom panel of
Figure~\ref{fig:fish2} is present in the original PostScript file.  And
every element in the original event display is preserved in the final
image.  Such fidelity is important;\footnote{\ldots but not universal
in scientific illustration.} the image is instantly readable to
physicists, including its creators in D\O, and intelligible---with just
a bit of explanation---to laypersons.  We wrote captions that would
explain and initiate conversation.  The postcard images
(\hhref{lutece.fnal.gov/Postcards}) have become true icons of particle
physics, with an impact far beyond their original purpose.

\subsection{Snowmass 2001}
The Division of Particles and Fields and the Division of Physics of
Beams of the American Physical Society organized a three-week summer
study on the future of particle physics in Snowmass, Colorado, in July
2001.  More than 1200 physicists participated---many young, many from
outside the United States---in a very broad examination of where our
field should be heading.  Early on, we decided to make outreach and
education an essential part of the Snowmass 2001 experience.  We wanted
to share our love for  science with the interesting mix of
people in Aspen and Snowmass, and to encourage our colleagues to see 
each other in action. We also believed that the public interaction 
would reinforce the optimism and enthusiasm that participants brought to 
Snowmass.

\begin{figure}[t]
   \begin{center}
       {\includegraphics[width=10cm]{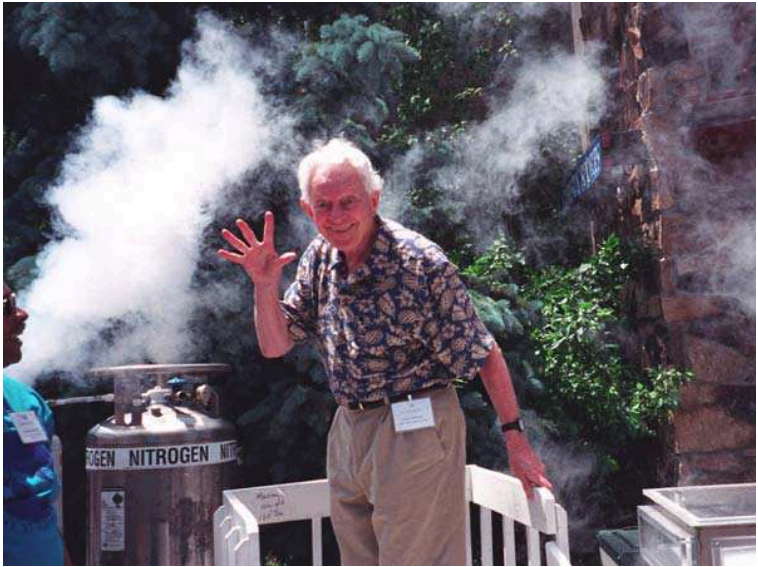}}
\caption{Leon Lederman levitated by the Meissner effect at Snowmass 
2001. (Photo credit: Elizabeth H.\ Simmons, Michigan State University)}
\label{fig:fish}
\end{center}
\end{figure}

Theoretical physicist Elizabeth Simmons, who chaired the outreach and
education effort, has described the extraordinary results
in \textit{Physics Today.}\cite{lizs} The Snowmass 2001 program
(\hhref{snowmass2001.org/outreach/education2.html}) included workshops
for teachers and students, public lectures, a science book fair,
science theater, open-air talks, astronomy activities, conversations
with children, outreach workshops, and a balloon ascension to recreate
 Victor Hess's discovery of cosmic rays.  The centerpiece was a huge
weekend science fair on the Snowmass Village Mall that attracted some
1500 members of the general public.  Among the weekend's hits 
was a superconducting apparatus capable of levitating an entire
human (see Figure~\ref{fig:fish}) from the Texas Center for
Superconductivity and Advanced Materials
(\href{http://www.tcsam.uh.edu/education_outreach}{www.tcsam.uh.edu/education\_outreach}).

\textit{Comunicare Fisica} participants will find special interest in 
the communications workshops 
(\hhref{www.fnal.gov/pub/snowmass/workshops/workshop.html}) organized 
at Snowmass 2001.

In connection with the summer study, the DPF commissioned an
illustrated thematic survey of our vision of particle physics and its
future in the most ambitious intellectual terms.  Within this broad
vision, the document was to identify the questions we want to address
over the next two decades.  Like the summer study itself, the thematic
survey aimed to help our community recognize and articulate what particle
physics is and aspires to be, guided by the scientific imperatives. 
\textit{Quarks Unbound} was a smashing success, for three principal 
reasons. First, we were careful to think through what we wanted to 
accomplish and to identify the audiences we wanted to reach. Second, 
we entrusted the project to a small team---not a large and 
representative committee---that included a science writer, a graphic 
designer, and a few physicists. [We did not want the sort of bland 
``offends no one, delights no one'' product that large committees are 
adept at producing!] Third, we distributed \textit{Quarks Unbound} 
widely and enthusiastically to groups and individuals. A private 
donor financed a second printing, so that the 
number of copies distributed worldwide now exceeds $60\,000$.

\section{Some lessons \label{sec:lessons}}
I am continually impressed by the passion, curiosity, and faith in the
value of exploration evidenced by those who attend our outreach and
education programs.  It may be true, as several speakers have opined,
that the general population is ignorant and indifferent about science,
but my experience is that we do find a receptive and engaged
audience.  Accordingly, the first lesson is: Respect (do not
underestimate) the audience.

Leadership from established scientists and Heroes of the Field 
provides important validation for the efforts of others. The 
effectiveness of leading by example is well-established. Respected 
senior physicists can also discourage the misperception that engaging 
with the public is unworthy of serious scientists. 
They also have value as fundraisers, from both public and 
private sources. 

While leaders are important, it may be even more important for the
leaders to let go: to grant autonomy and resources to small groups and
trust them to do wonderful things.  Nothing is more stifling to
creativity and innovation than repeated reviews by committees
constructed to represent an average.  The field is better served by
original---even quirky---efforts that explore a whole range of
approaches than by efforts programmed to hit the mean each time.  Both
those in authority and those who execute should find pleasure in
experimenting and taking risks!


A familiar conceit is that ``Physicists can do anything.''  I like to
tell my colleagues that the complete slogan concludes ``\ldots badly,''
to remind them of the rewards of collaborating with professionals who
know their fields as well as we physicists know physics.  Working with
gifted writers, editors, designers, artists, and educators can be immensely
satisfying and can lead you to do things you didn't know were possible.

Let me underline again the importance of local activities.  These
include organizing public lectures and symposia at the universities and
labs, engaging with cultural institutions and creative people within
the community, and developing relationships with local radio stations
and newspapers.  A record of success on the local scene may even enable
effective national efforts.  

One committed person can do a lot---and an individual can achieve even
more with encouragement and support from colleagues. But it is also 
delightful to learn from others and to draw inspiration from their 
efforts. The \textsl{QuarkNet} model has succeeded so well because it has 
elements of individuality, collaboration, coordination, and the bond 
that grows from being part of a large and ambitious enterprise.

Always ask, Why are we doing this?  Who is the audience?  What are
our goals?  What will success mean?  Be prepared to finish the task:
if you prepare a new brochure about your experiment or your subject and
are passive or hesitant about distributing it, you have limited your effectiveness.
If following through seems like too much effort, you should notice that
before you start.  Expect to discover new people (not all PhD
physicists).  Your role may be catalytic.  Leaders succeed when others
become the stars.  Expect also to discover new ideas, and remember that
a new outreach triumph may occur when you least expect it.

\section{Concluding remarks}
Experiments at the Large Hadron Collider break new ground in scale and
internationalism, counting collaborators in the thousands.  The
outreach efforts of the ATLAS (34 countries) and CMS (31 countries)
collaborations are truly transnational, and marvelously dynamic.  The
ATLAS experiment movie (\hhref{atlasexperiment.org/movie/}), for
example, is available in ten languages and has won awards in five
countries.

\begin{figure}[t!]
   \begin{center}
       {\includegraphics[width=11cm]{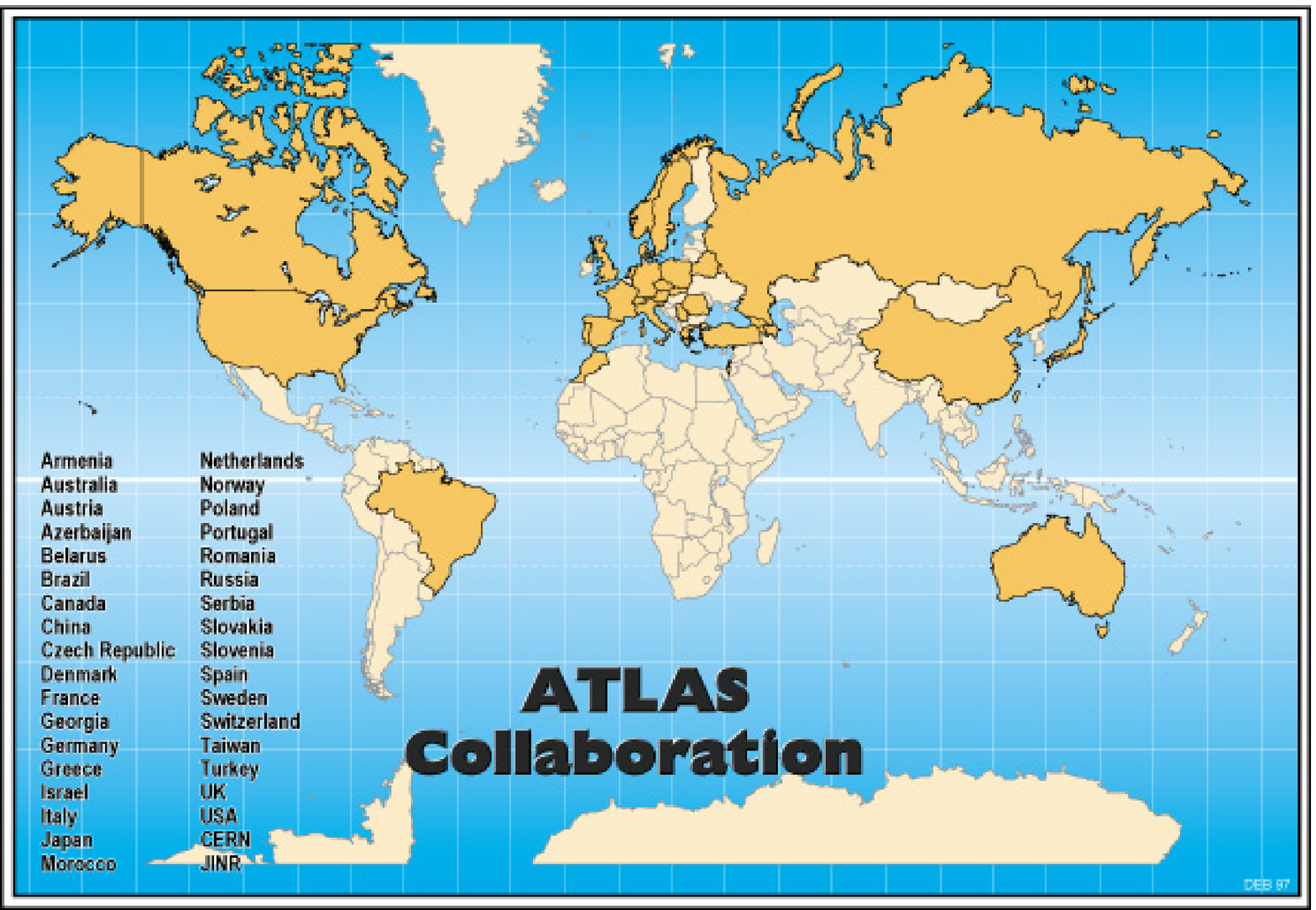}}\\[3pt]
       {\includegraphics[width=11cm]{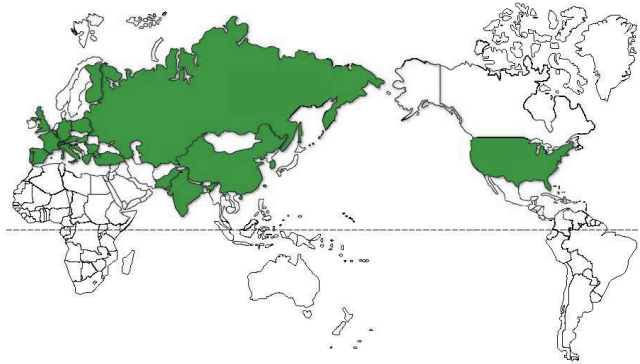}}
\caption{Global reach of the ATLAS and CMS collaborations at the LHC.}
\label{fig:cms}
\end{center}
\end{figure}

The impressive reach of ATLAS and CMS, depicted by the colored spaces in
Figure~\ref{fig:cms}, is a source of immense satisfaction and pleasure to
particle physicists.  For me, and for many of my colleagues, the
opportunity to meet and join in common cause with people from many
nations is one of the great joys of our research.  But there are many
blank spots on the LHC maps of the world, and we should take those open
spaces as a challenge.  I believe that we should aspire to engage the
whole world in the values and the rewards of science.  Our goal should
truly be physics without boundaries.

Science is more than solving the next great puzzle of particle physics;
it is a set of values for contemplating the world.  I draw great hope
from the concluding lines of Anthony Lewis's millennial essay in the
New York \textit{Times} of December 31, 1999. You will recognize the 
tradition we owe to Galileo:
\begin{quote}
    [T]here has been one transforming change over
    this thousand years. It is the adoption of the
    scientific method: the commitment to
    experiment, to test every hypothesis. But it is
    broader than science. It is the open mind, the
    willingness in all aspects of life to consider
    possibilities other than the received truth. It is
    openness to reason. 
\end{quote}
When we are at our best---when we are truest to these ideals---we do our  best 
science, and we give our greatest gift to society. 

\section{Acknowledgements}
I am grateful to Franco Fabbri and Rinaldo Baldini Ferroli for their
kind invitation to participate in \textit{Comunicare Fisica 2005,} and
for generous hospitality in Frascati.  I thank Marge Bardeen, Michael
Barnett, Sharon Butler, Julia Child, Aaron Dominguez, Judy Jackson,
Leon Lederman, Joe Lykken, Kate Metropolis, Helen Quinn, Liz Quigg,
Randy Ruchti, Liz Simmons, Maria Spiropulu, and  other
colleagues for teaching me many ways to communicate physics.  Fermilab
is operated by Universities Research Association Inc.\ under Contract
No.\ DE-AC02-76CH03000 with the U.S.\ Department of Energy.

\section*{Appendix: Some education and outreach resources \label{app}}

{\scriptsize
{ \centering
\begin{tabular}{rl} 
    High-Energy Physics Outreach  &
    \hhref{www-ed.fnal.gov/hep/} \\[3pt]
    Fermilab's \textit{Saturday Morning Physics} & 
    \hhref{www-ppd.fnal.gov/smp-w/}\\
    Argonne National Lab & \hhref{www.anl.gov/Careers/Education/}\\
    Berkeley Lab & \hhref{csee.lbl.gov/} \\
    Brookhaven National Lab & \hhref{www.bnl.gov/scied/}\\
    Cornell Lab for Elementary Particle Physics & 
    \hhref{www.lns.cornell.edu/public/outreach/}\\
    Kavli Institute for Cosmological Physics & \hhref{cfcp.uchicago.edu/education/}\\[3pt]
    Illinois Physics Van & \hhref{van.hep.uiuc.edu} \\
    Maryland Physics is Phun & 
    \hhref{www.physics.umd.edu/PhysPhun/}\\
    Michigan State Science Theatre & \hhref{www.sciencetheatre.org}\\
    Little Shop of Physics & 
    \hhref{littleshop.physics.colostate.edu} \\
    UCSB Physics Circus & \href{http://www.physics.ucsb.edu/~circus/}{www.physics.ucsb.edu/\~{}circus/} 
    \\[3pt]
    CPEP &
    \hhref{www.cpepweb.org}\\[3pt]
    The Particle Adventure & \hhref{particleadventure.org}\\[3pt]
    \textsl{QuarkNet} & \hhref{quarknet.fnal.gov }\\[3pt]
    \textit{Understanding the Universe} & \hhref{www-ed.fnal.gov/uueo/} \\[3pt]
    \textsl{QuarkNet Grid} & \hhref{quarknet.fnal.gov/grid} \\[3pt]
    \textit{Mariachi} & \href{http://www.phy.bnl.gov/~takai/MariachiWeb/}{www.phy.bnl.gov/\~{}takai/MariachiWeb/} \\[3pt]
    \textit{CHEPREO} & \hhref{www.chepreo.org} \\[3pt]
    \textit{NALTA} & \hhref{csr.phys.ualberta.ca/nalta}\\[3pt]
    \textit{CROP} & \href{http://cse.unl.edu/~gsnow/crop/crop.html}{cse.unl.edu/\~{}gsnow/crop/crop.html} \\[3pt]
    Web Lecture Archives & \hhref{www-visualmedia.fnal.gov} \\
     & \hhref{webcast.cern.ch/Projects/WebLectureArchive/} \\
     & \hhref{www-project.slac.stanford.edu/streaming-media/}\\
      & \hhref{www.itp.ucsb.edu/talks/}\\[3pt]
     Tevatron Postcards & \hhref{lutece.fnal.gov/Postcards}
     \\[3pt]
     \textit{Quarks Unbound} & \href{http://www.aps.org/units/dpf/quarks_unbound/}{www.aps.org/units/dpf/quarks\_unbound/} \\[3pt]
    LIGO & \hhref{www.ligo-la.caltech.edu/public.htm} \\
     & \hhref{www.ligo-wa.caltech.edu} \\
     & \hhref{www.einsteinathome.org} \\[3pt]
     LHC-CMS & \hhref{uscms.fnal.gov} \\
      & \hhref{cmsinfo.cern.ch/Welcome.html} \\[3pt]
      LHC-ATLAS & \hhref{atlasexperiment.org} \\[3pt]
      Research experiences  & \hhref{www.nsf.gov/crssprgm/reu/index.jsp} \\[3pt]
      World Year of Physics & \hhref{www.physics2005.org} \\ 
\end{tabular}
}
}

\vspace*{12pt}
\noindent
The page \hhref{particleadventure.org/particleadventure/other/othersites.html} 
contains a very extensive list of internet resources.
\end{document}